\documentclass[review]{elsarticle}

\usepackage{lineno,hyperref}
\usepackage{amsmath}

\journal{Journal of Crystal Growth}

\makeatletter
\newcommand{\figcaption}[1]{\def\@captype{figure}\caption{#1}}
\newcommand{\tblcaption}[1]{\def\@captype{table}\caption{#1}}
\makeatother

\usepackage[hang,small,bf]{caption}
\usepackage[subrefformat=parens]{subcaption}
\usepackage{braket}

\begin{document}

\begin{frontmatter}

\title{Shape-controlled crystal growth of Y$_3$Al$_5$O$_{12}$:Ce single crystals with application of micro-pulling-down method and Mo crucibles, and their scintillation properties}

\author[address1]{Masao Yoshino\corref{mycorrespondingauthor}}
\cortext[mycorrespondingauthor]{Corresponding author}
\ead{yoshino.masao@tohoku.ac.jp}

\author[address1]{Atsushi Kotaki}
\author[address1]{Yuui Yokota}
\author[address1]{Takahiko Horiai}
\author[address1,address2,address3]{Akira Yoshikawa}

\address[address1]{Institute for Material Research, Tohoku University, Miyagi 980-8577, Japan}
\address[address2]{New Industry Creation Hatchery Center, Tohoku University, Sendai, Miyagi 980-8579, Japan}
\address[address3]{C\&A Corporation, Sendai, Miyagi, 980-0811, Japan}




\begin{abstract}
The technology to grow single crystals of the required shape directly from a melt has been researched extensively and developed in various industries and research fields. In this study, a micro-pulling-down method and a Mo crucible were applied to the shape-controlled crystal growth of Y$_3$Al$_5$O$_{12}$:Ce (YAG:Ce).
Three types of Mo crucibles with different die shapes were developed.
Stable crystal growth in the same shape as the die was achieved, and scintillation properties that are comparable with those of the previously reported YAG:Ce were obtained.
\end{abstract}

\begin{keyword}
A2. Micro-pulling-down method \sep 
A2. Shape-controlled crystal growth \sep 
B2. Scintillator materials \sep
B3. Scintillators \sep 
B1. Oxides
\MSC[2010] 00-01\sep  99-00
\end{keyword}

\end{frontmatter}


\section{Introduction} \label{sec:intro}
Inorganic scintillator have been applied as radiation imaging sensors in many fields \cite{Nikl:2015bc0,shimazoe2020,kodama2020}. Some studies have developed imaging techniques that use scintillator arrays as collimators \cite{lee2014,omata2020}.
A tube-shaped single-crystal scintillator was required in these studies to achieve a high image quality.
Scintillators that are used in medical applications are typically cut and polished from a large bulk single-crystal form and assembled into radiation detectors. The Czochralski \cite{kamada2012,kochu2019}, Bridgman--Stockbarger \cite{yoshikawa2016}, and Floating Zone \cite{yokota2007} methods are commonly used as crystal-growth techniques to obtain large bulk crystals.
However, it is difficult to process large single-crystal scintillators into a tube, which has been an obstacle to the production of tube-shaped scintillators.

The technology to grow single crystals of the required shape directly from the melt has been researched extensively and developed in various industries and research fields. Edge-defined film-fed growth (EFG) \cite{kamada2020,kurlov1998} and micro-pulling-down ($\mu$--PD) methods \cite{yoshikawa2009,yoshikawa2004,yoshikawa2007} are examples of such technology.
The $\mu$--PD method is a crystal growth technique that uses a crucible with a die with several holes in the bottom to supply the melt. 
It is possible to control the crystal shape with the same cross-sectional shape as the die. 
In our recent research, a tubular Y$_3$Al$_5$O$_{12}$:Ce (YAG:Ce) single-crystal scintillator was grown by the $\mu$--PD method and Ir crucible \cite{kotaki2020}.
The rising price of Ir crucibles is a major problem in the mass production of single crystals. As an alternative to precious metals such as Ir, crystal growth that uses inexpensive crucibles such as Mo is promising, and the growth of YAG:Ce single crystals using Mo crucibles has been reported recently \cite{Tkachenko2018}.

In this study, the $\mu$--PD method and Mo crucibles with different die shapes were applied to growing tube shaped-controlled YAG:Ce (tube-YAG:Ce) single-crystal scintillators. The optical and scintillation properties of the grown crystals were evaluated.

\section{Materials and Methods} \label{sec:mat}
\subsection{Mo crucible with a tube-shaped die} \label{ssec:crucible}
\begin{figure}[ht]
\begin{minipage}[t]{0.9\linewidth}
 \centering
 \includegraphics[keepaspectratio, width=0.6\linewidth]{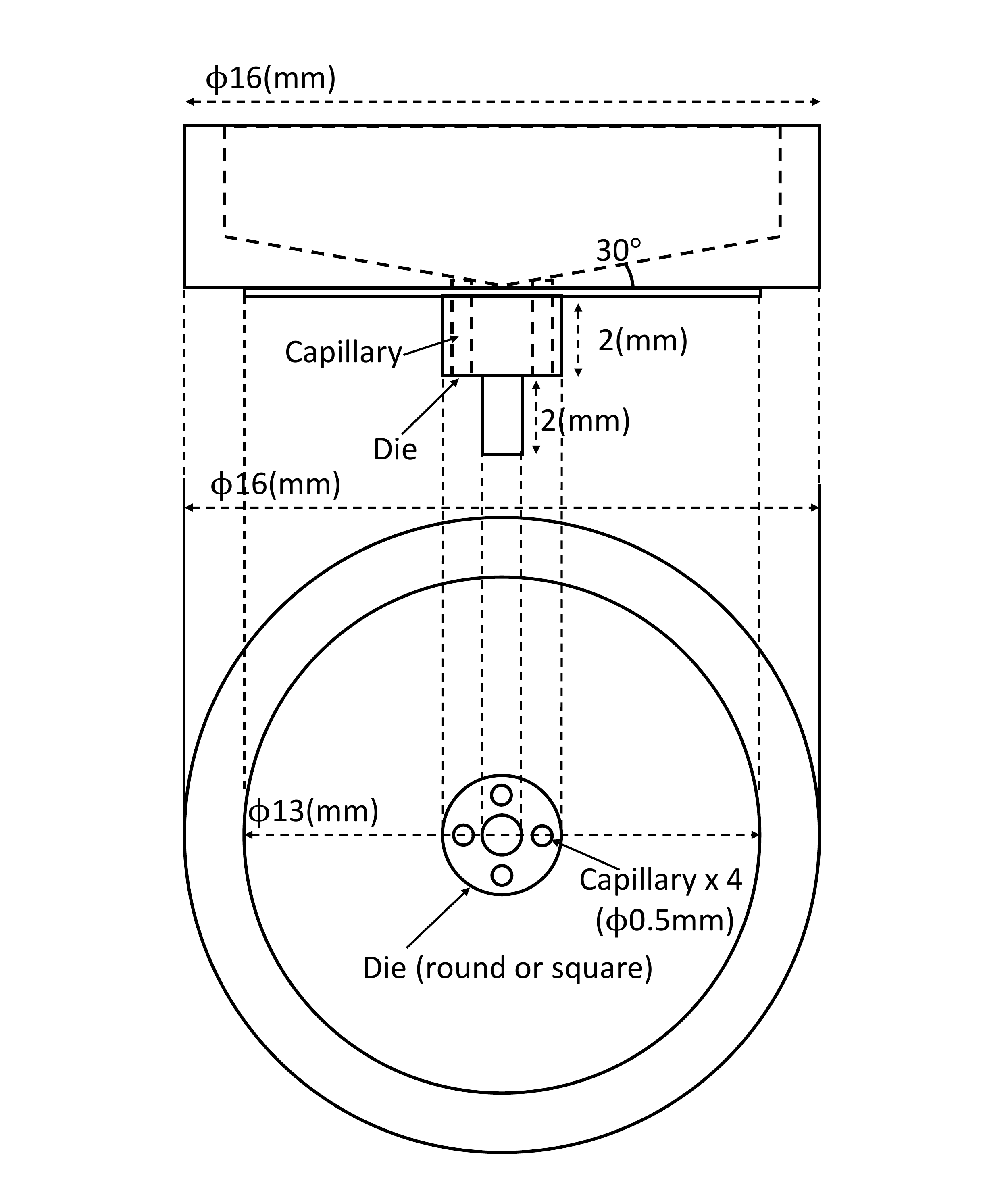}
 \subcaption{}
 \label{subfig:dim_crucible}
\end{minipage}\\

\begin{minipage}[ht]{0.9\linewidth}
 \centering
 \includegraphics[keepaspectratio, width=0.9\linewidth]{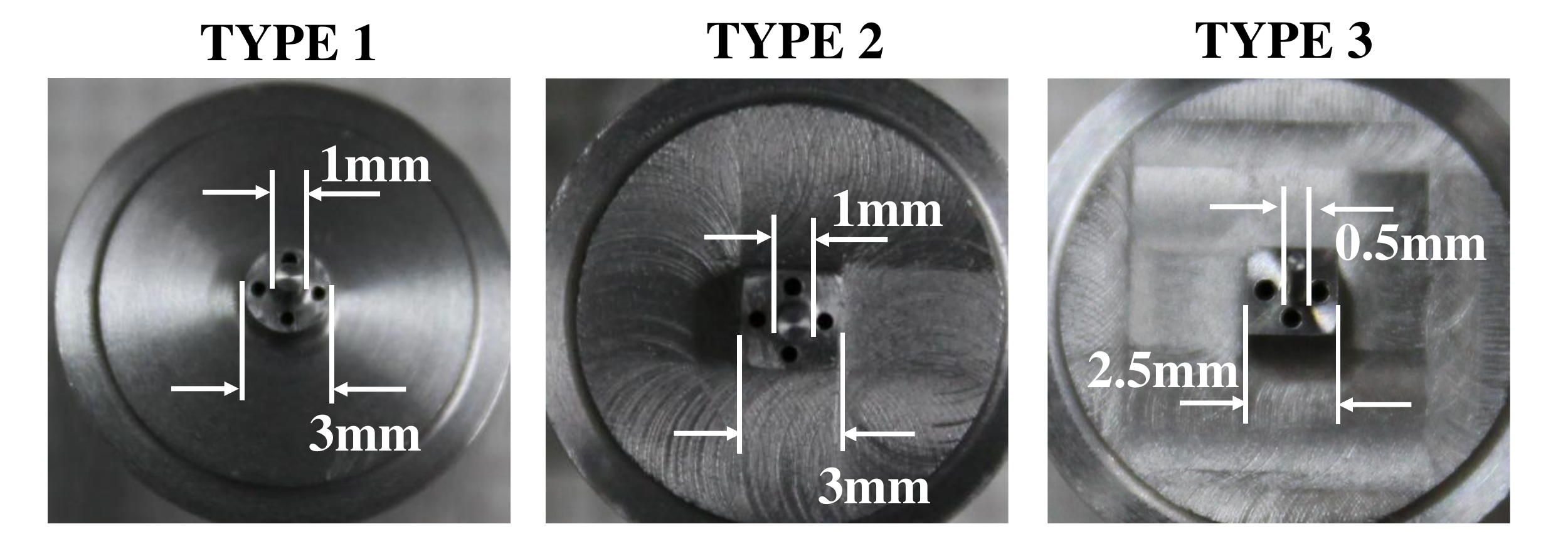}
 \subcaption{}
 \label{subfig:photo_crucible}
\end{minipage}
 \caption[dimension crucible]{Design drawing and photographs of developed Mo crucibles.}
 \label{fig:crucible}
\end{figure}

The design drawings of the developed Mo crucible are shown in Fig. \ref{subfig:dim_crucible}.
We developed three types of Mo crucibles with different die shapes (Fig. \ref{subfig:photo_crucible}). 
The TYPE1 crucible has a round die shape, and TYPE2 and TYPE3 crucibles have a square die shape. By using these crucibles, we investigated the effect of die shape on crystal growth. 

\subsection{Crystal growth setup} \label{ssec:crystal}
\begin{figure}[ht]
 \centering
 \includegraphics[keepaspectratio, width=.8\linewidth]{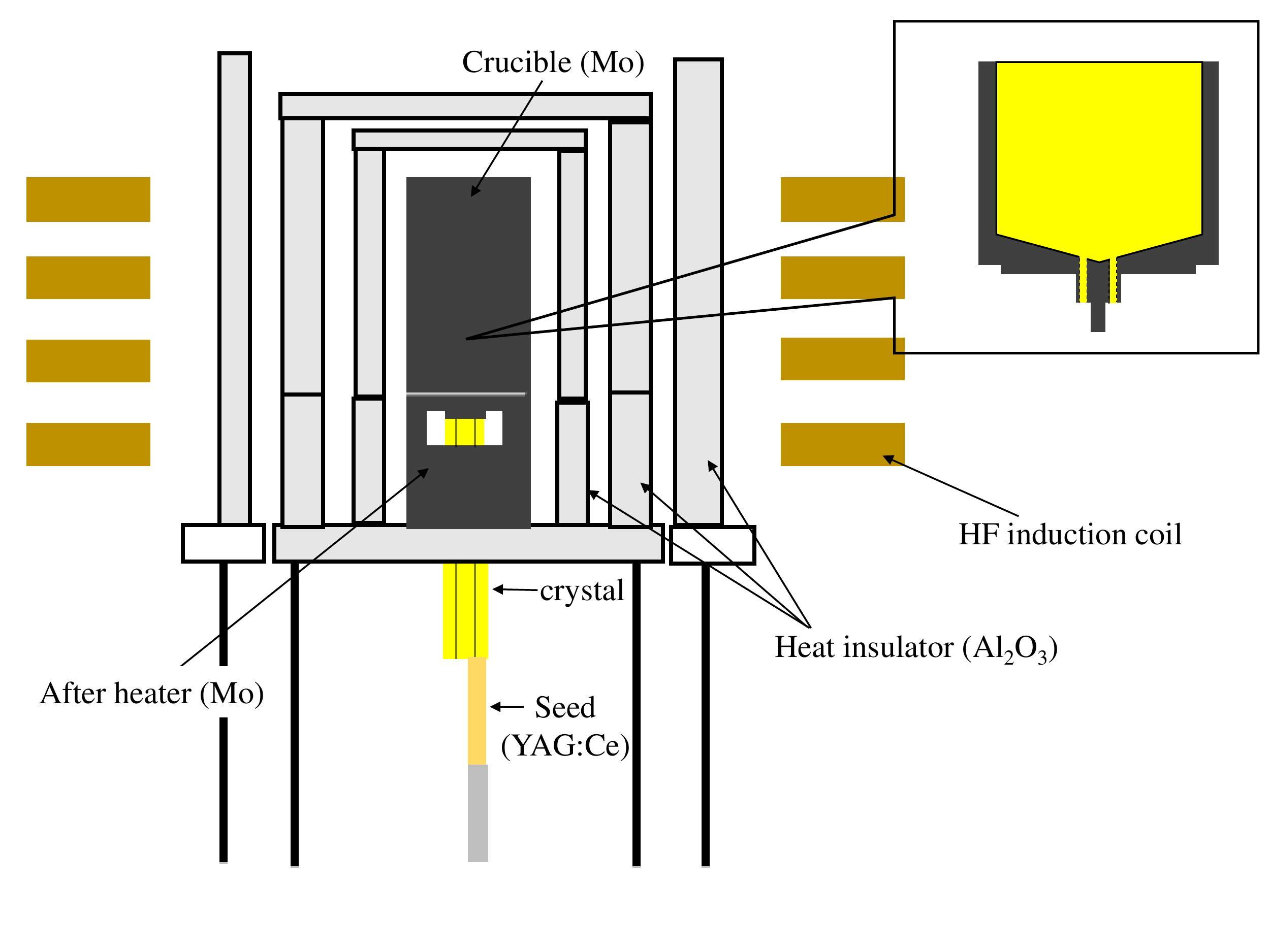}
 \caption[u-PD schematic]{Schematic diagram of crystal-growth assembly.}
 \label{fig:u-PD}
\end{figure}

The $\mu$-PD method was applied to grow tube-YAG:Ce single crystals.
Powder starting materials $\alpha$-Al$_2$O$_3$, Y$_2$O$_3$, and CeO$_2$ (purity 4N or higher) were weighed out and mixed to obtain nominal compositions of (Y$_{0.995}$Ce$_{0.005}$)$_3$Al$_5$O$_{12}$. The mixed powder was placed in a crucible and heated to the melting point of YAG:Ce by using a high-frequency induction coil.
An Ar + H$_2$ atmosphere was used to prevent the Mo crucibles from oxidizing during crystal growth. The seed crystals were $\braket{0~0~1}$-oriented YAG:Ce 1.0\% single crystals. 
A schematic diagram of the $\mu$-PD crystal growth assembly is provided in Fig. \ref{fig:u-PD}. The growth rate varied from 0.1 mm/min to 0.5 mm/min to grow tube-shaped crystals.
The grown tube-YAG:Ce single crystals were cut perpendicular to the growth direction and polished to produce evaluation samples for characterization.

\subsection{Characterization} \label{ssec:chara}
The chemical composition of the polished sample surface was analysed by using an electron probe microanalyser (EPMA, JXA-8530F, JEOL Ltd., Tokyo, Japan) to evaluate the radial Ce distribution.
Transmittance spectra were measured from 200-750~nm by using a UV/VIS spectrophotometer (V-550, JASCO Corp., Tokyo, Japan).
X-ray irradiated luminescence spectra were measured at room temperature by using an X-ray tube system (Mini-X, AMPTEK Inc., Boston, MA, USA, 20 kV, 150$~\mu$A). The luminescence was detected by a Shamrock 163 Spectrograph (Andor technology Ltd., Belfast, North Ireland) coupled with a Newton DU420-OE CCD camera (Andor technology Ltd., Belfast, North Ireland).
Scintillation lights were collected by a photomultiplier tube (PMT, R7600-200, Hamamatsu Photonics K.K., Hamamatsu, Japan) with an operating voltage of 600 V. A 10~MBq $^{137}$Cs source was used at a distance of 5 cm. The signal from the PMT was amplified by a charge-sensitive pre-amplifier (113, ORTEC, Oak Ridge, TN, USA), shaping amplifier (572A, ORTEC, Oak Ridge, TN, USA), and digitized by a multi-channel analyser (Pocket MCA 8000A, AMPTEK Inc., Boston, MA, USA). 
The scintillation waveforms were recorded by using a digital oscilloscope (TDS3035B, Tektronix, Inc., Beaverton, OR, USA). The decay time was calculated by fitting one or a sum of two exponential functions to the waveform:

\begin{displaymath}
 A(t) = A_1 exp(-t/\tau_1) + A_2 exp(-t/\tau_2) + const
\end{displaymath}
The values $A_1$,$A_2$ and $\tau_1$,$\tau_2$ are the amplitudes and decay times of the waveforms, respectively.

\section{Results and Discussion}
\label{sec:result}
\subsection{Crystal growth}
\begin{figure}[ht]
\begin{minipage}[t]{0.45\linewidth}
\centering
\includegraphics[keepaspectratio, scale=0.39]{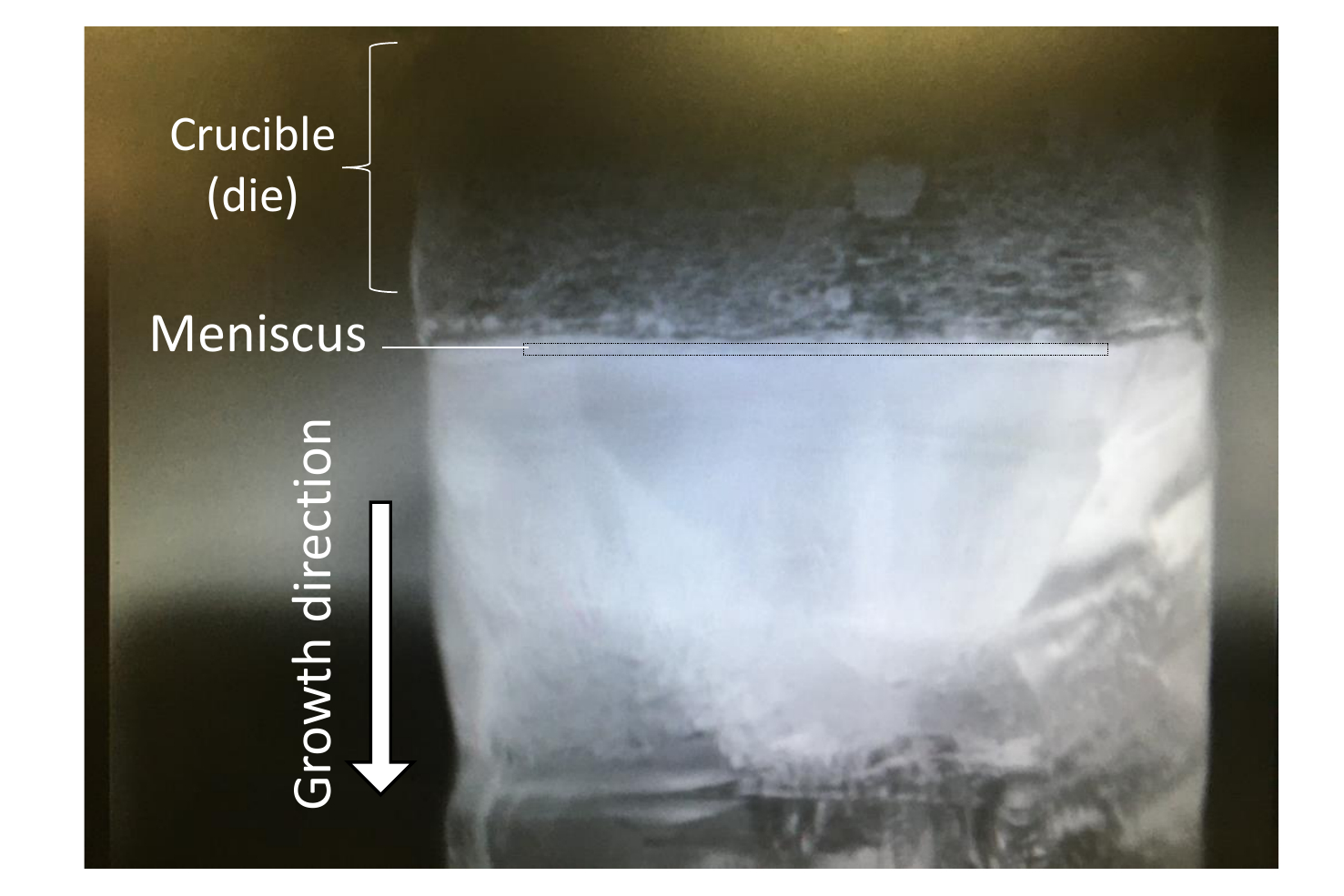}
\subcaption{Early stage growth}
\label{subfig:before}
\end{minipage} ~
\begin{minipage}[t]{0.45\linewidth}
\centering
\includegraphics[keepaspectratio, scale=0.41]{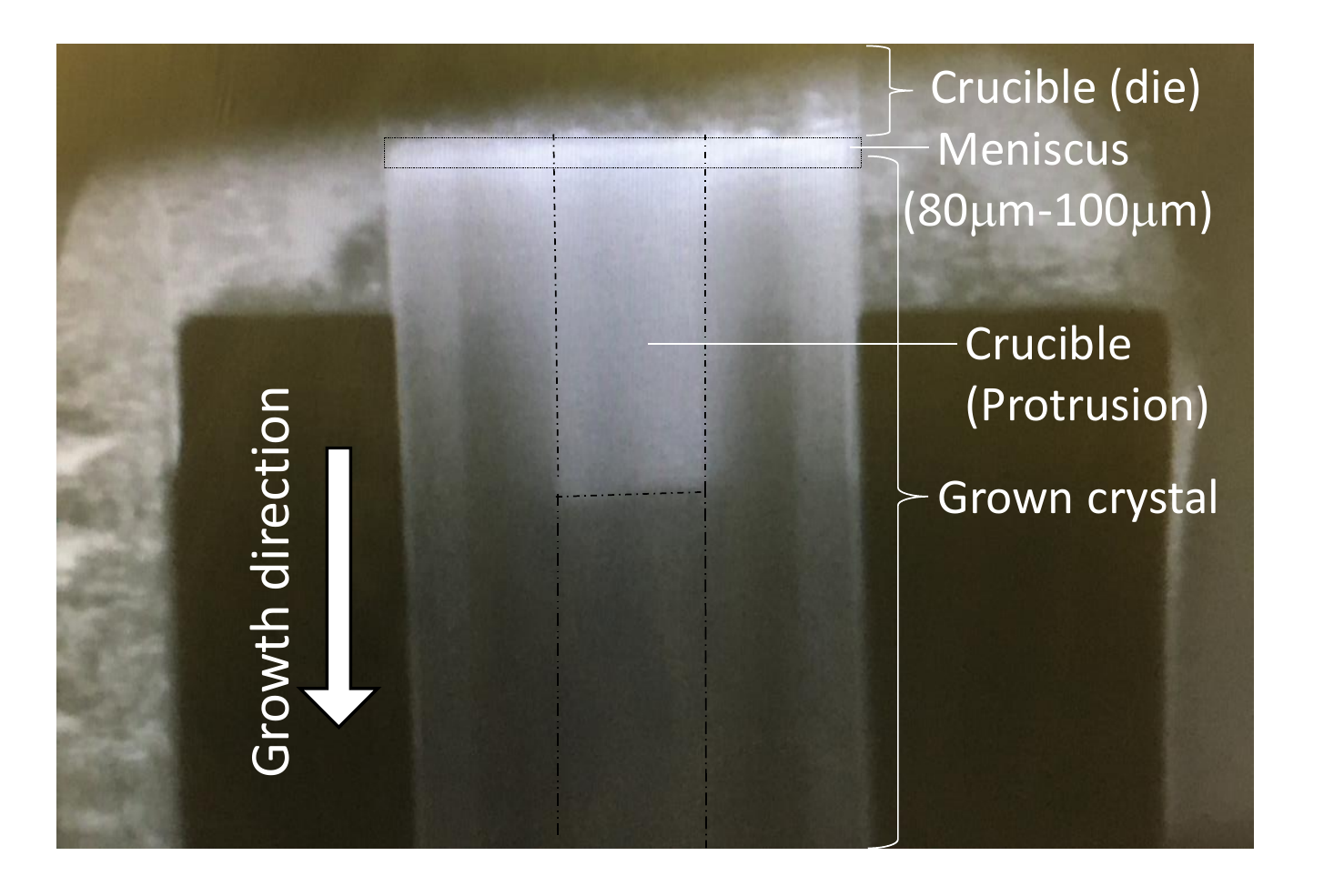}
\subcaption{After stabilization of meniscus}
\label{subfig:after}
\end{minipage} \\
\begin{minipage}[t]{0.9\linewidth}
\centering
\includegraphics[keepaspectratio, scale=0.5]{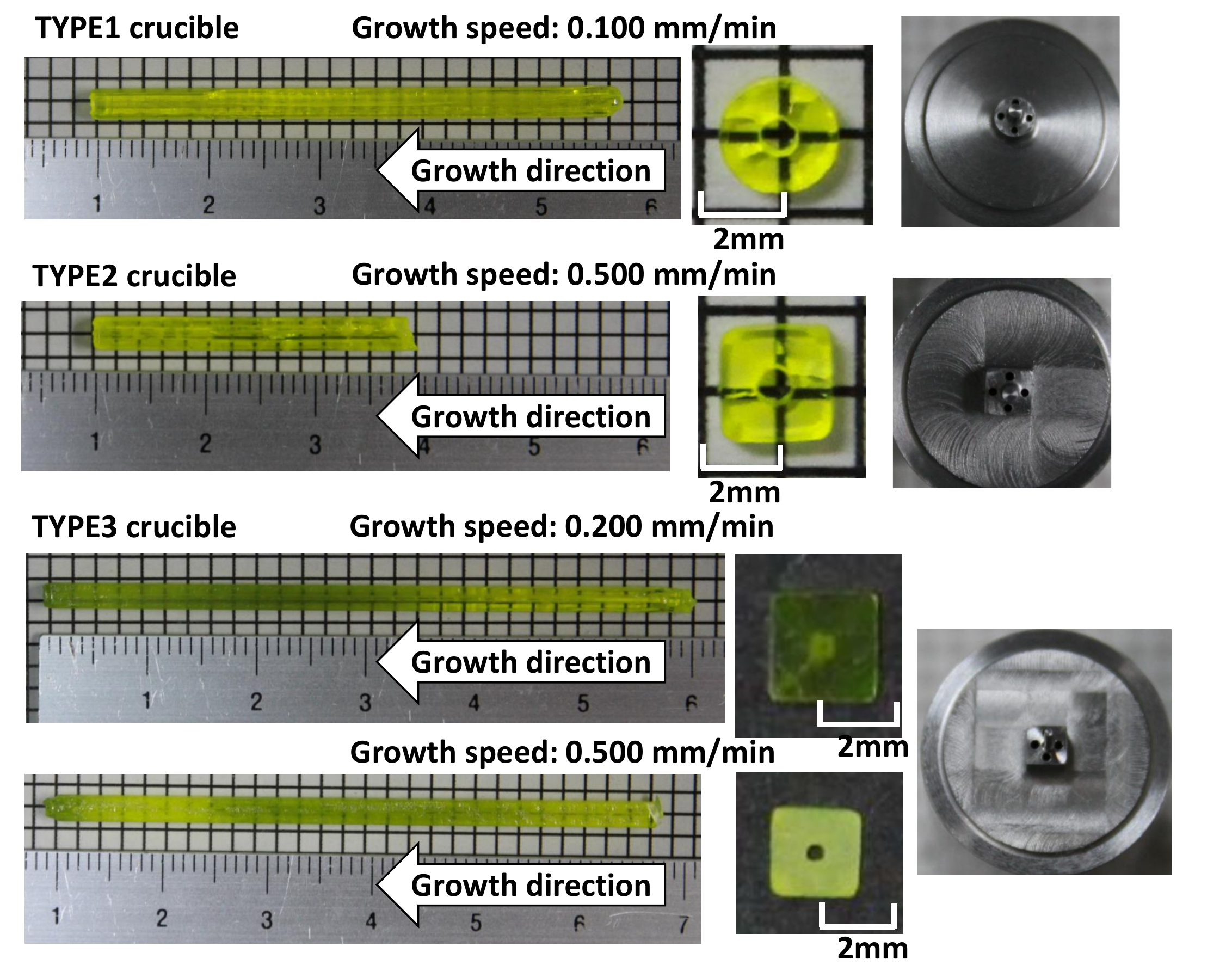}
\subcaption{Grown tube-shaped crystals.}
\label{subfig:crystals}
\end{minipage}    
\caption{(a), (b)Liquid--solid interface during crystal growth and (c) tube-shaped crystals grown by three types of Mo crucibles.}
\label{fig:interface}
\end{figure}

\begin{figure}[ht]
 \centering
 \includegraphics[keepaspectratio, width=\linewidth]{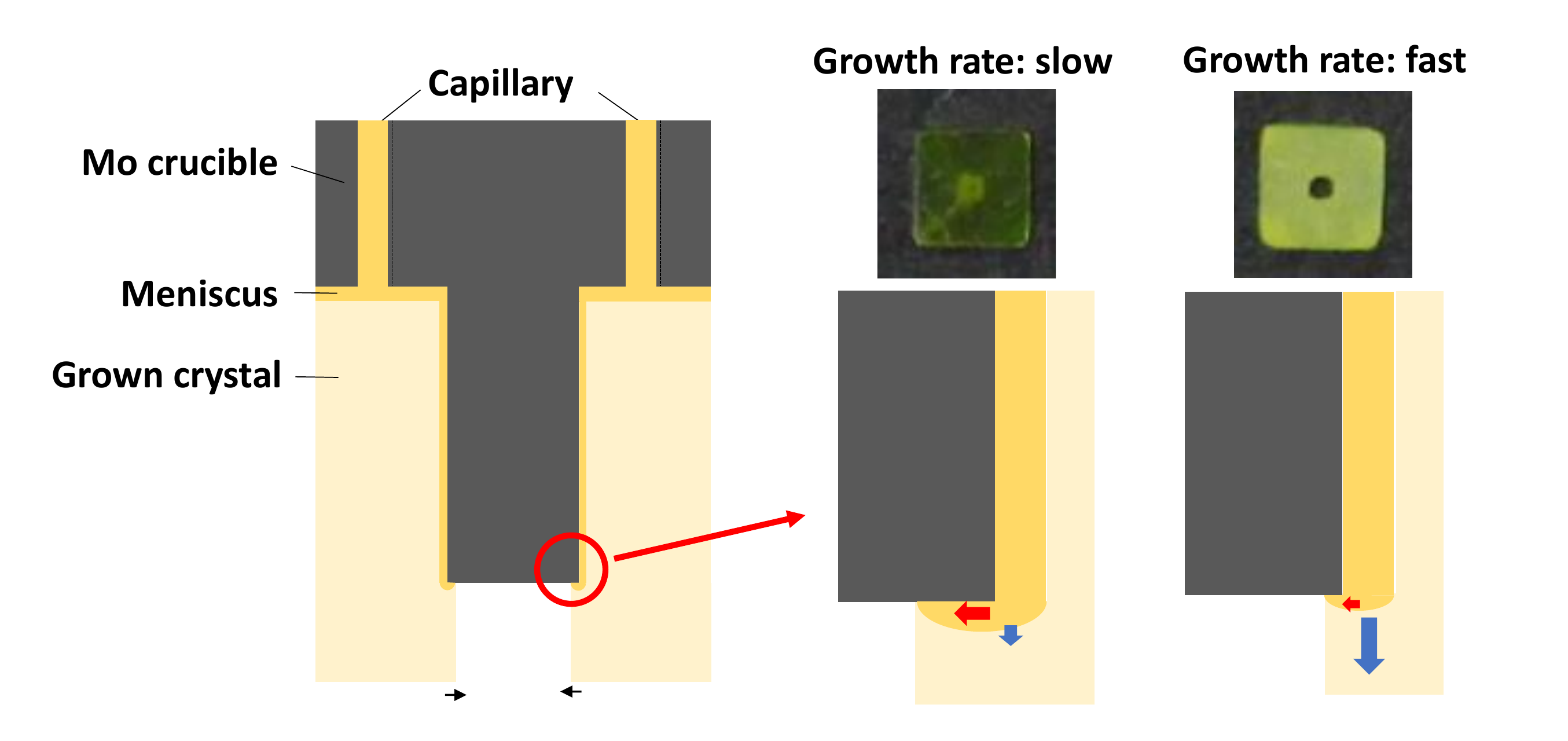}
 \caption{Schematic diagram of effect of growth speed on control of internal crystal diameter}
 \label{fig:growth_mechanism}
\end{figure}

The crystal growth of the tube-YAG:Ce single crystals was achieved by the $\mu$-PD method.
After the starting material had been melted in a Mo crucible, the melt passed through capillaries to reach the die at the bottom of the crucible. The seed crystal was touched to the melt at the bottom of the die to form a meniscus between the bottom of the die and the seed crystal. The crystals were grown continuously by pulling down the seed crystal with a moderate temperature gradient near the liquid--solid interface (Fig. \ref{fig:interface}).
In the early stage of crystal growth (Fig. \ref{subfig:before}), the spreading of the meniscus to the bottom of the crucible was unstable, which resulted in an inconsistent outer diameter of the grown crystals. Figure \ref{subfig:after} shows the solid--liquid interface when the meniscus thickness was stabilised by optimising the power of the high-frequency (HF) induction coil and the pulling-down rate.
Photographs of the grown crystals and samples that were cut perpendicular to the growth direction are shown in Fig. \ref{subfig:crystals}. The cross-sectional shape of the grown crystals was like the shape of the bottom of the die. 
A comparison of the crystal growth in the TYPE1 and TYPE2 crucible shows the effect of die shape on crystal growth as a function of pulling-down speed for stable crystal growth. The pulling-down speed to stabilize crystal growth was faster in the crucible with a square die (0.5 mm/min) than in the crucible with a round die (0.1 mm/min).
Growth using the TYPE3 crucible was conducted at 0.2 mm/min and 0.5 mm/min. 
At a slow growth rate (0.2 mm/min), the meniscus was wetted and spread at the tip of the protrusion, and the inside of the tube crystal was filled with melt.
At high growth rates (0.5 mm/min), the inside of the tube crystal was not filled with melt because the wetting spread of the meniscus to the tip of the protrusion was suppressed (Fig. \ref{fig:growth_mechanism}).

\begin{figure}[ht]
\begin{minipage}[t]{0.32\linewidth}
\centering
\includegraphics[keepaspectratio, scale=0.45]{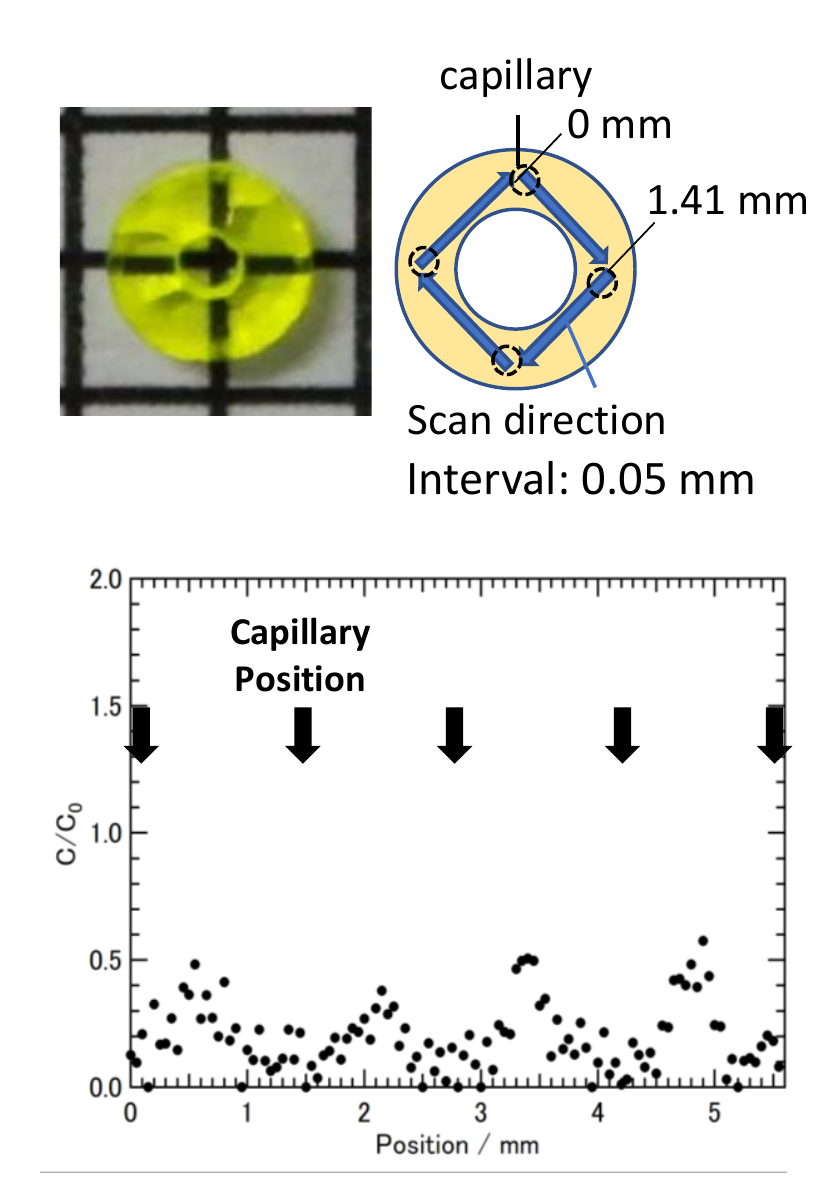}
\subcaption{TYPE1}
\label{subfig:epma_type1}
\end{minipage} 
\begin{minipage}[t]{0.32\linewidth}
\centering
\includegraphics[keepaspectratio, scale=0.45]{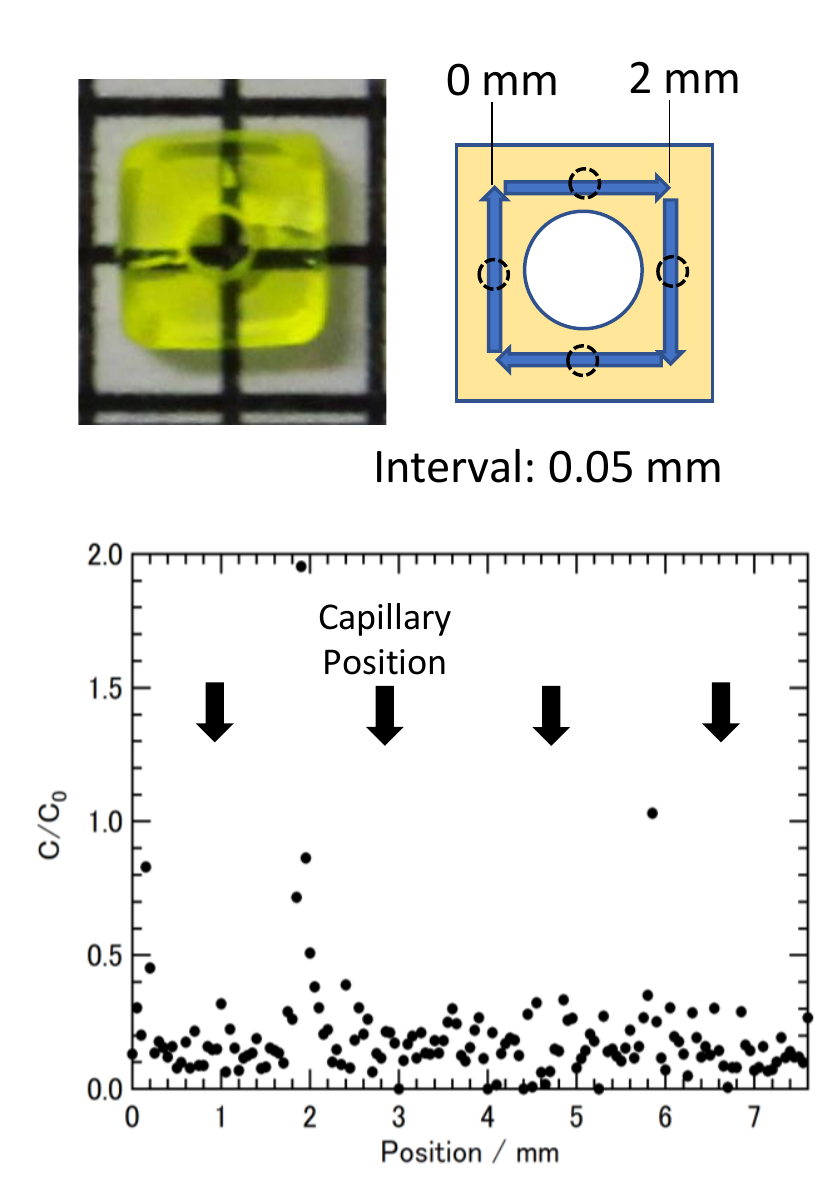}
\subcaption{TYPE2}
\label{subfig:epma_type2}
\end{minipage} 
\begin{minipage}[t]{0.32\linewidth}
\centering
\includegraphics[keepaspectratio, scale=0.45]{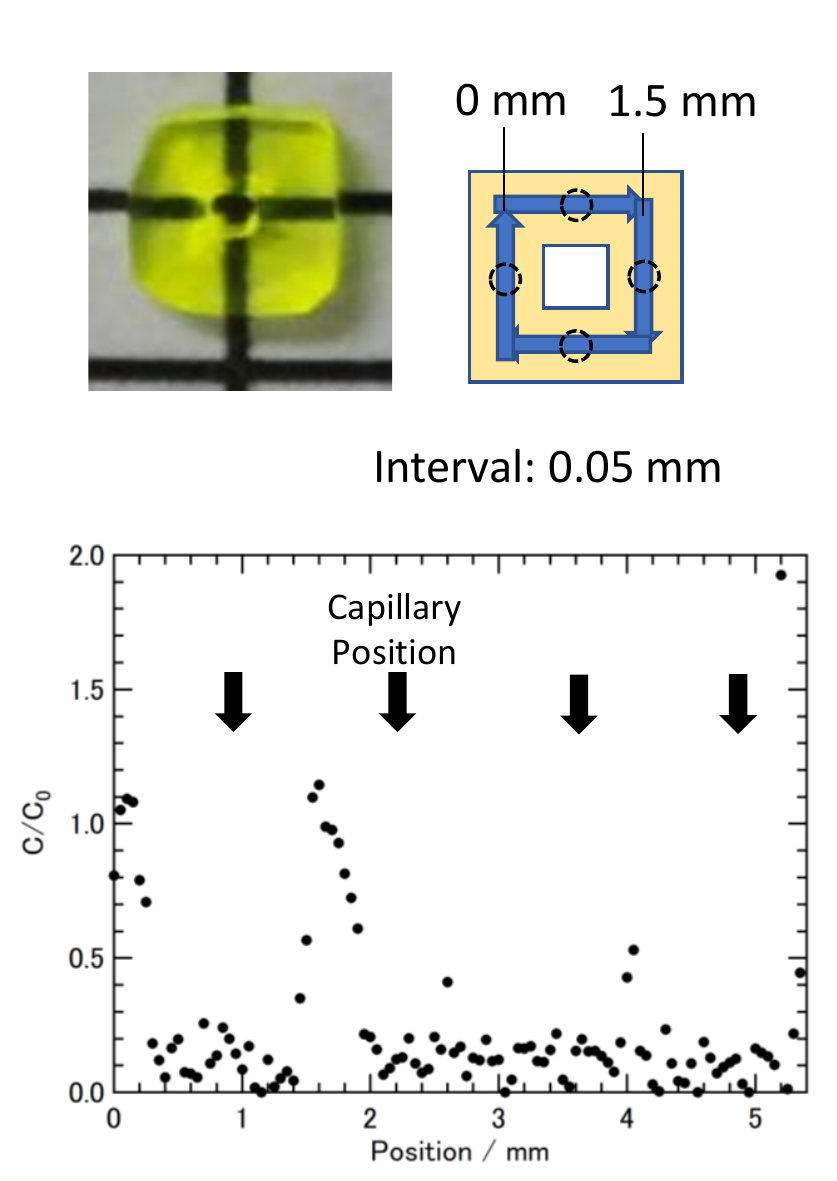}
\subcaption{TYPE3}
\label{subfig:epma_type3}
\end{minipage}    
\caption{Ce distribution in tangential direction of grown crystals.}
\label{fig:epma}
\end{figure}

Ce distribution in the tangential direction of the grown crystals was measured by EPMA (Fig. \ref{fig:epma}). The Ce concentration around the capillary tended to be smaller than that in the remainder of the capillary area.
This result agrees with our previous report on the growth of tube-YAG:Ce using Ir crucibles \cite{kotaki2020} -- the small Ce concentration around capillaries was caused by the small segregation coefficient of Ce in the YAG single crystal and by small convection in the meniscus.

\subsection{Optical and scintillation properties}
\begin{figure}[ht]
 \centering
 \includegraphics[keepaspectratio, width=\linewidth]{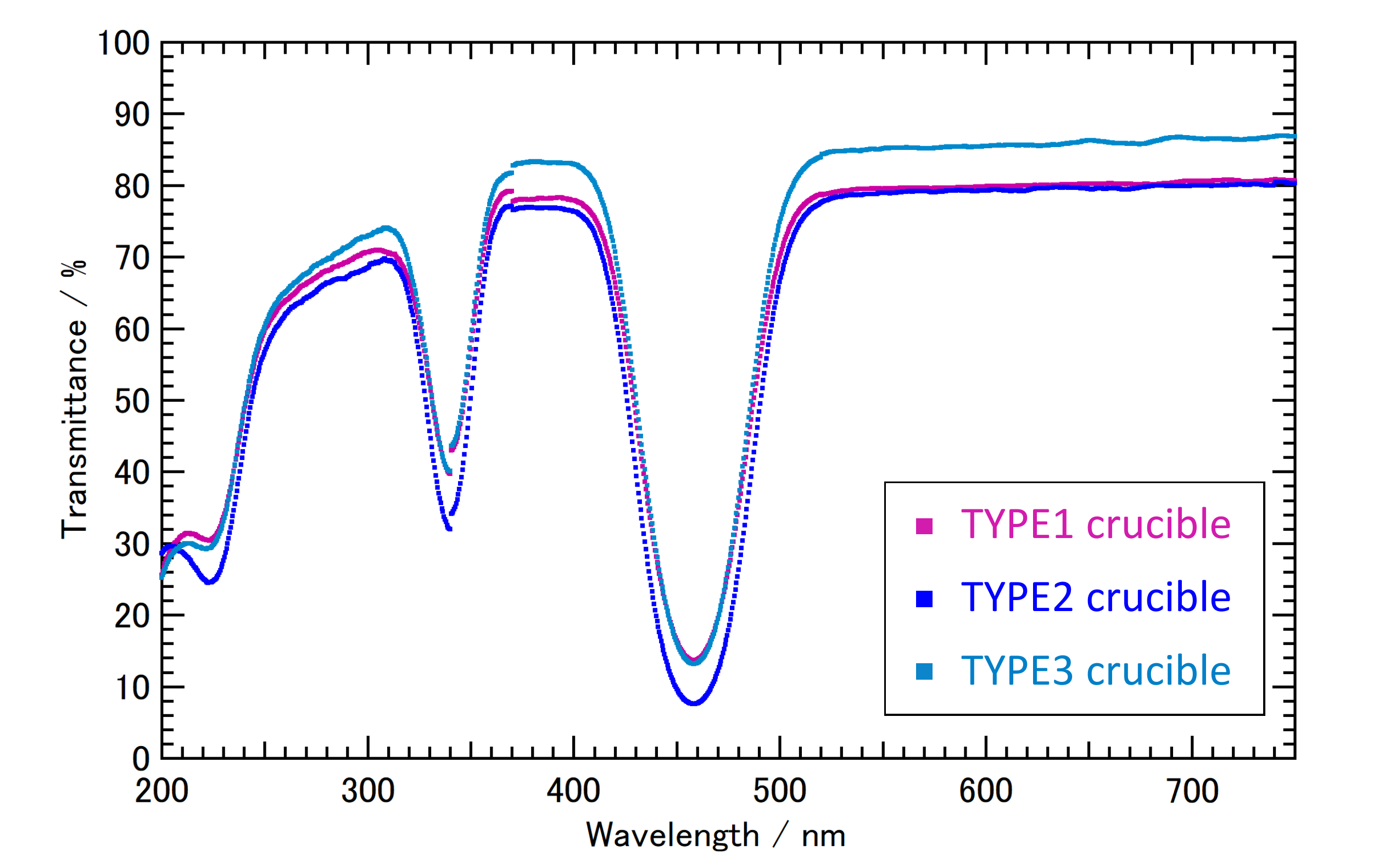}
 \caption{Optical transmittance spectra of grown crystals.}
 \label{fig:transmittance}
\end{figure}

The optical transmittance spectra of the tube-YAG:Ce crystals that were grown in this study are shown in Fig. \ref{fig:transmittance}. For all samples, two clear absorption peaks existed at 340~nm and 455~nm, which are ascribed to 4f $\rightarrow$ 5d$_2$ and 4f $\rightarrow$ 5d$_1$ absorption transitions, respectively. The transmittance above 500 nm exceeded 70\% for all samples.

\begin{figure}[ht]
 \centering
 \includegraphics[keepaspectratio, width=\linewidth]{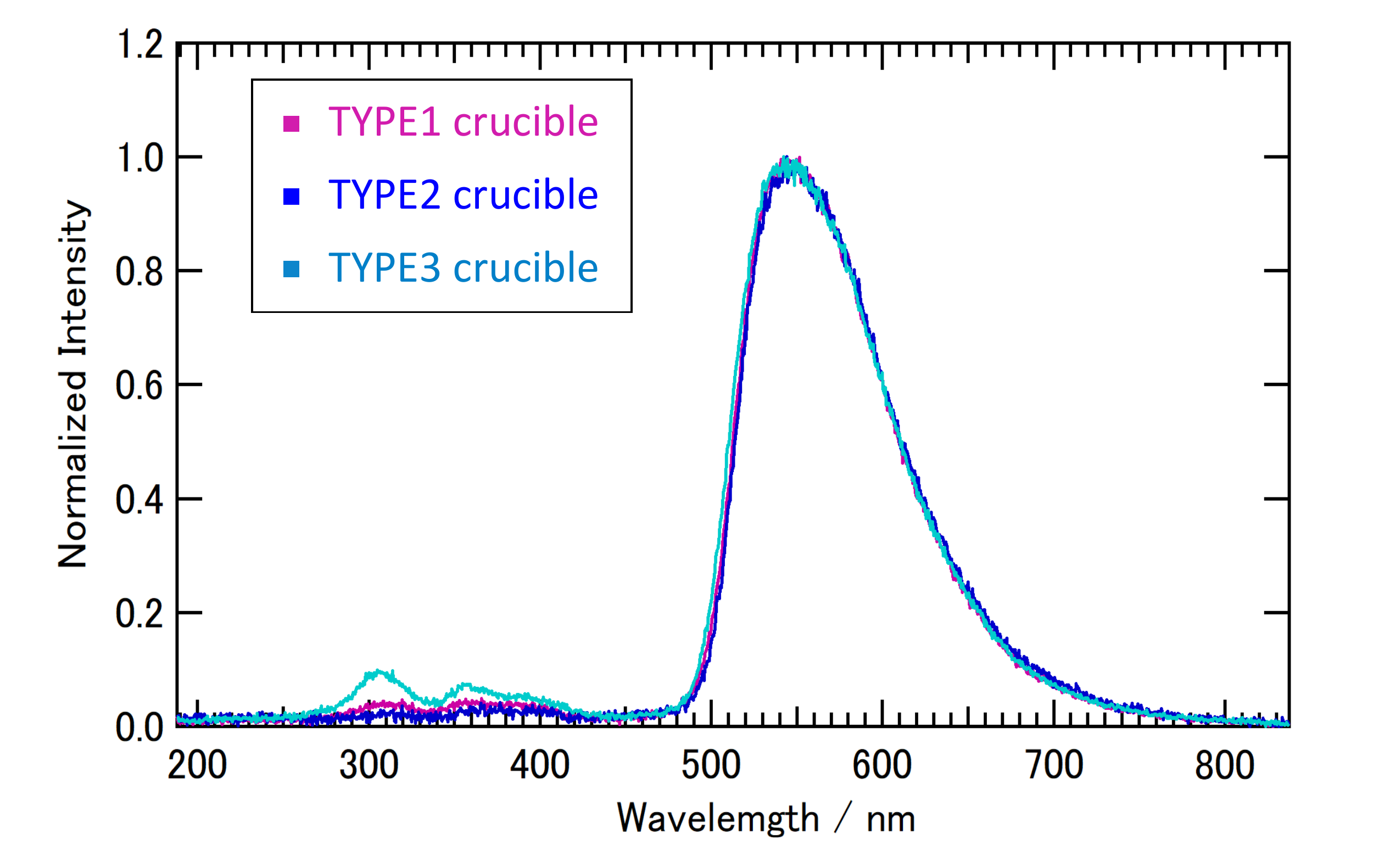}
 \caption{Radioluminescence spectra of grown crystals.}
 \label{fig:radilumi}
\end{figure}

The radioluminescence spectra of grown tube-YAG:Ce crystals are shown in Fig. \ref{fig:radilumi}. All sample shapes were the same with a broad-band emission and peak at 540 nm, which is typical for Ce$^{3+}$ 5d-4f emission in YAG crystals.
The small and broad emission around 250--400 nm is ascribed to self-trapped exciton and YAl anti-site defects, and a possible trap candidate has been reported to be antisite Y$^{3+}_\mathrm{Al}$ defects \cite{babin2005}.

\begin{figure}[ht]
 \centering
 \includegraphics[keepaspectratio, width=\linewidth]{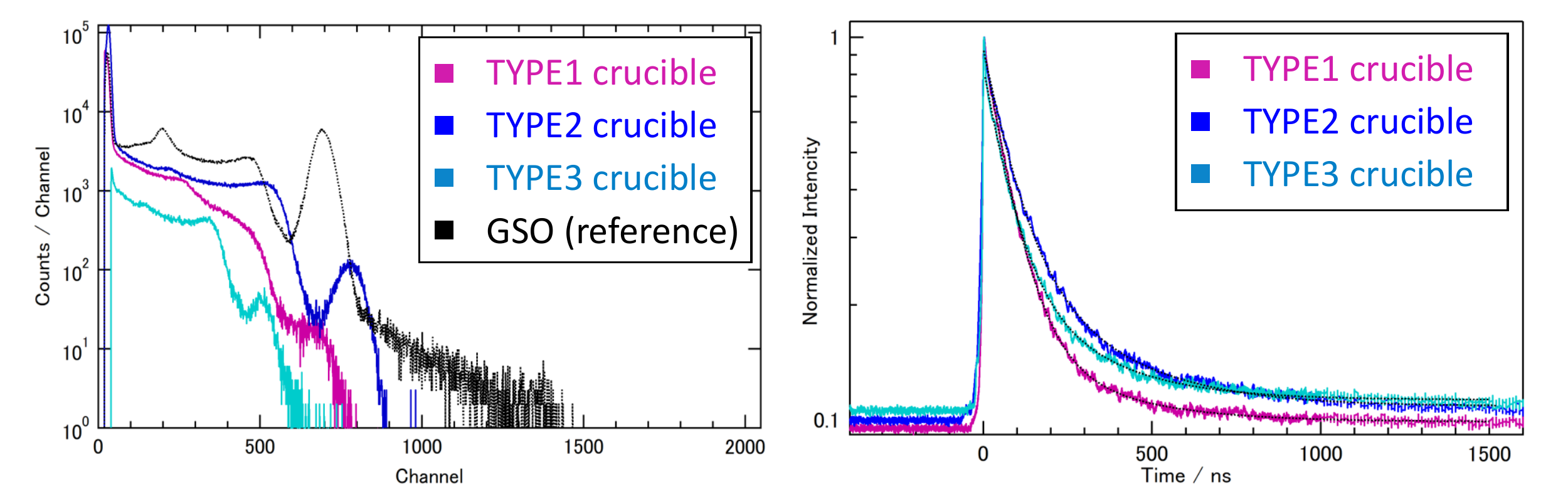}
 \caption{Scintillation pulse height spectra and decay curve of crystals grown by Mo crucibles.}
 \label{fig:scinti}
\end{figure}

Scintillation pulse height spectra and decay curves of tube-YAG:Ce that was grown by Mo crucibles were measured (Fig. \ref{fig:scinti}). Three samples were prepared by changing the cut-out position of the sample. The scintillation properties are summarised in Table \ref{tb:scint}. In the sample that was grown in the TYPE2 crucible, the scintillation properties were comparable with those reported previously \cite{moszynski1994}.

\begin{table}[ht]
  \centering
  \caption{Results of scintillation property measurements}
  \begin{tabular}{lcrr}
    Position    &   Light output        &   Primary decay   &   Secondary decay \\ \hline
    TYPE1   &   19000 ph/MeV   &   71 ns   &   260 ns   \\
    TYPE2   &   23000 ph/MeV  &   77 ns   &   260 ns   \\
    TYPE3   &   15000 ph/MeV   &   67 ns   &   230 ns  \\ \hline
  \end{tabular}
  \label{tb:scint}
\end{table}

\section{Conclusion}
\label{sec:conc}
Three types of Mo crucibles with different die shapes were developed.
The crucibles were applied to grow tube-YAG:Ce single crystals by using the $\mu$--PD method.
In the early stages of crystal growth, meniscus spreading into the die was unstable, and the outer diameter of the grown crystals fluctuated.
After the meniscus thickness stabilized, the diameter of the grown crystals became uniform, and the cross-sectional shape of the crystals was close to that of the die.
The optimum pulling-down speed for crystal-growth stabilization was faster in the crucible with a square die than that in the crucible with a round die.
When grown in a crucible with a small inner diameter (TYPE 3), at a slow growth rate (0.2 mm/min), the meniscus was wetted and spread at the tip of the protrusion, and the inside of the tubular crystal was filled with melt. However, an increase in growth rate to 0.5 mm/min suppressed the filling of the interior of the tube crystal.
Measurements of the tangential Ce concentration distribution in the grown crystals showed that the Ce concentration around the capillary was lower than in the remainder of the crystal.
The transmittance of grown crystals above 500 nm was almost the same and above 70\%.
The scintillation properties of the tube-YAP:Ce single crystals that were grown in the TYPE2 crucible were comparable with those of previously reported YAG:Ce single crystals.

\section*{Acknowledgement}
This work was supported by JSPS KAKENHI [grant numbers 21H03834, 19K17191].
We would like to thank Mr. Narita for his support about EPMA measurement.
We also thank Laura Kuhar, PhD, from Edanz (https://jp.edanz.com/ac) for editing a draft of this manuscript.

\bibliography{mybibfile}

\end{document}